\documentclass[journal,comsoc]{IEEEtran}
\usepackage[T1]{fontenc}% optional T1 font encoding
\usepackage{amsmath}
\usepackage[cmintegrals]{newtxmath}
\usepackage{graphicx}
\usepackage{acro}
\usepackage{todonotes}
\usepackage{url}

\DeclareAcronym{IETF}{
short = IETF ,
long = Internet Engineering Task Force
}
\DeclareAcronym{NGA}{
short = NGA ,
long = Next Generation Access
}
\DeclareAcronym{FCC}{
short = FCC ,
long = Federal Communications Commission
}
\DeclareAcronym{BBF}{
short = BBF ,
long = Broadband Forum
}
\DeclareAcronym{CPE}{
short = CPE ,
long = Customer Premises Equipment
}
\DeclareAcronym{HCPE}{
short = HCPE ,
long = Hybrid CPE
}
\DeclareAcronym{HAG}{
short = HAG ,
long = Hybrid Access Gateway
}
\DeclareAcronym{DSLAM}{
short = DSLAM ,
long = Digital subscriber Line Access Multiplexer
}
\DeclareAcronym{GRE}{
short = GRE ,
long = Generic Routing Encapsulation
}
\DeclareAcronym{PE}{
short = PE ,
long = Provider Edge
}
\DeclareAcronym{LTE}{
short = LTE ,
long = Long Term Evolution
}
\DeclareAcronym{LAG}{
short = LAG ,
long = Link Aggregration Group
}
\DeclareAcronym{VLANs}{
short = VLANs ,
long = Virtual Local Area Networks
}
\DeclareAcronym{DSL}{
short = DSL ,
long = Digital Subscriber Line
}
\DeclareAcronym{MPTCP}{
short = MPTCP ,
long = Multipath TCP
}
\begin{document}

\title{Increasing broadband reach with \\Hybrid Access Networks
}
\author{\IEEEauthorblockN{Nicolas Keukeleire, Benjamin Hesmans, Olivier Bonaventure}\\
\IEEEauthorblockA{Tessares, Louvain-la-Neuve, Belgium\\
Email: \texttt{First.Last@tessares.net}}}

\maketitle

\begin{abstract}
End-users and governments force network operators to deploy faster Internet access services everywhere. Access technologies such as FTTx, VDSL2, DOCSIS3.0 can provide such services in cities. However, it is not cost-effective for network operators to deploy them in less densely populated regions. The recently proposed Hybrid Access Networks allow to boost xDSL networks by using the available capacity in existing LTE networks.  We first present the three architectures defined by the Broadband Forum for such Hybrid Access Networks. Then we describe our experience with the implementation and the deployment of Multipath TCP-based Hybrid Access Networks.

\end{abstract}


\section{Introduction}

\IEEEPARstart{N}{etwork} operators have deployed a variety of fixed access network technologies (xDSL, DOCSIS, FTTx, \ldots) and various wireless technologies (3G, 4G, FWA and soon 5G). Today's bandwidth hungry applications like video streaming and various cloud services have forced the network operators to increase the capacity of their access networks. During the last decades, various improvements have been brought in \ac{DSL} networks. While the initial deployments offered bandwidths of only a few Mbps, current deployments with VDSL2 and G.Fast have reached hundreds of Mbps or more. Both VDSL2 and G.Fast can reach their peak bandwidth when the households are close to the street cabinet (from hundreds of meters for G.Fast to a kilometre for VDSL2). These technologies are effective in dense areas where fiber has been deployed close to the end-users. However, there are many regions, such as rural areas, where xDSL cannot reach its optimal bandwidth due to signal loss on long copper pairs.

Governments, in Europe, America and Asia have announced ambitious objectives to provide better broadband services to their entire population. In Europe, the objective is to provide 30 Mbps to all Europeans by 2020. %\footnote{\url{https://ec.europa.eu/digital-single-market/en/policies/broadband-europe}}. 
In the USA, broadband is defined as 25 Mbps in downstream and 3 Mbps in upstream. %\footnote{\url{https://www.fcc.gov/reports-research/reports/broadband-progress-reports/2018-broadband-deployment-report}}.

The European Commission defines the \ac{NGA} as \emph{fixed-line broadband access technologies capable of achieving download speeds meeting the Digital Agenda objective of at least 30 Mbps coverage} \cite{ECreport}. Based on data from the 28 countries of the European Union, DSL is the most widespread non-wireless technology in rural areas with a coverage of 86.3\% in 2017, but it is not considered as an \ac{NGA}. VDSL is only available in 32.5\% of the rural areas. FTTP, Cable and DOCSIS barely reach 10\% of coverage in these areas. On the other hand, \ac{LTE} covers 89.9\% of the rural areas across Europe in 2017. The latest measurements reported by the \ac{FCC} in the USA \cite{fcc} indicate that only 92.3\% of the population has access to a fixed terrestrial service at 25 Mbps. At 50 Mbps, this number drops to 90.8\% of the population. 

%\todo{To be reviewed} 
%\begin{figure}[htbp]
%    \center
%    \includegraphics[width=0.4\textwidth]{figures/EU28.png}
%    \caption{Source: Broadband Coverage in Europe 2017, a study by IHS Markit and Point Topic for the European Commission}
%    \label{fig:EU28}
%\end{figure}

To cope with these government objectives, network operators deploy \ac{NGA} technologies in densely populated areas such as cities. However, this deployment takes time and will still require several years to complete. In rural areas, \ac{NGA} technologies are barely economically viable given the population density. This has encouraged network operators to explore alternatives. One of them is to combine a terrestrial broadband access network such as xDSL with a wireless access network such as \ac{LTE}. %Wireless networks are usually deployed to support smartphones and mobile users. Measurements indicate that lunch-time is usually the peak time for mobile networks while xDSL's peak times usually occur in the evening. This implies that the mobile network has unused capacity during the peak times of the xDSL network.
The combined availability of xDSL and LTE opens new opportunities for the network operators that own these two types of networks. The Hybrid Access Networks discussed in this paper make it possible to combine the existing xDSL network with an existing wireless network such as \ac{LTE} to provide higher bandwidth services. This enables network operators to provide faster Internet access services without the costly fiber roll-outs that are required by solutions such as VDSL2 or G.Fast. 

During the last few years, Hybrid Access Networks have moved from lab prototypes to large-scale deployments of standardized solutions. Standards are important for network operators because they allow solutions from different vendors to inter-operate. We first describe in Section~\ref{section:standards} the reference architectures and the key protocols that the \ac{BBF} and the \ac{IETF} have specified to build Hybrid Access Networks. Then, Section~\ref{section:implem} analyses measurements of the key components of such networks and provides feedback on real deployments.

\section{Hybrid Access Networks standards}\label{section:standards}

Network operators rely on different standardized technologies to offer broadband services to their subscribers. Standards have became a prerequisite for large scale deployments of access network solutions by network operators. The \ac{BBF} initiated architectural work that was finalized in 2016 \cite{bbf348} while the work on the protocols was progressing within the \ac{IETF}. The \ac{BBF} is currently finalizing the \emph{Nodal Requirements for Hybrid Access Broadband Networks} \cite{bbf378} that leverage recent \ac{IETF} specifications \cite{rfc6824,rfc8157,draft-converters}. 

\subsection{Hybrid Access Architecture}
To understand how the Hybrid Access Networks will be deployed, it is important to first understand their architecture. A classical access network includes subscriber's clients using Wi-Fi or Ethernet connected to a \ac{CPE}. The latter is then connected to the operator's network through one of technologies discussed in the previous sections, which provides Internet access. In a typical xDSL deployment, the \ac{CPE} receives an IP adress from the DSL provider. For simplicity, we consider in this section that addresses are allocated to the \ac{CPE} by the network without discussing the differences between IPv4 and IPv6. We use the \texttt{@} symbol to represent such an address. %It then uses one \texttt{/64} prefix for the local LAN and address \texttt{p::c/64} is assigned to the client. 
If the \ac{CPE} is attached to a single xDSL network, then all the Internet packets to/from the client are transmitted over the xDSL link. Some operators have deployed bonded DSL solutions where two DSL lines are terminated on the same \ac{CPE}. This provides higher bandwidth to end-users, but network operators report that it is difficult to deploy such solutions at a large scale because they typically require the two bonded links to be terminated on the same card on the same \ac{DSLAM}. Given these operational problems, many network operators do not consider bonding two DSL lines as a cost-effective solution to provide higher bandwidth services.

%Due to space limitations, we consider that only IPv6 is used, even though hybrid access networks support both IPv6 and IPv4. Given the scarcity of IPv4 addresses, IPv4 solutions need to include Network Address Translation functions that add some complexity.  

\begin{figure}[ht]
     \centering
     \includegraphics[width=0.45\textwidth]{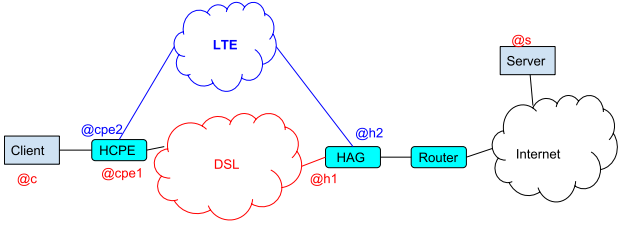}
     \caption{The architecture of Hybrid Access Networks}
     \label{fig:architecture}
 \end{figure}

Now, let us consider what happens when the \ac{CPE} is attached to both an xDSL and a \ac{LTE} network as in Figure~\ref{fig:architecture}. We call such a \ac{CPE} a \ac{HCPE} in this paper. From an addressing viewpoint, there are two possible architectures to connect these \ac{HCPE}s to the network. A first approach is to configure the address allocation function in \textbf{both} the xDSL and the \ac{LTE} network to allocate the same address on both links (i.e. addresses \texttt{@cpe1} and \texttt{@cpe2} in Figure~\ref{fig:architecture} are equal). This requires a strong coordination between the address allocation services on both the xDSL and the \ac{LTE} network. In this case, the HCPE is reachable through the same address over the two networks. It can load-balance the packets that it sends over the two access links at any time and the network can also load balance the packets over the two access links. Several load-balancing schemes are possible on the HCPE and in the network. With per-flow load balancing, each UDP flow or TCP connection is associated to one access link and all the packets belonging to this flow are sent over this link. This solution works well, but it implies that a single flow can only use the bandwidth of one access link. If a single flow needs to be able to utilize the bandwidth of the two access links, then the HCPE and the network must use a per-packet load balancing strategy. With such a strategy, different packets belonging to a single TCP connection can be sent over different access networks. However, xDSL and \ac{LTE} networks experience different delays. This implies that the packets sent over the two networks will be reordered before reaching their final destination. TCP reacts to such reordering by retransmitting packets and reducing its congestion window which in the end slows down the data transfer. Thus, although the TCP packets that belong to a given TCP connection can be sent over both the xDSL and the 4G network, in practice doing so results in reduced performance due to TCP collapse. 

From an address allocation viewpoint, the simplest solution to connect the HCPE to both the xDSL and the \ac{LTE} network is to allocate a different address on each access link. An important point to note about this address allocation is that network providers usually implement \emph{Network Ingress Filtering} \cite{rfc2827} to prevent spoofing attacks. The consequence of Network Ingress Filtering is that if address \texttt{@a1} (resp. \texttt{@a2}) has been assigned on the xDSL link (resp. the \ac{LTE} interface), then only the packets whose source address is equal to \texttt{@a1} (resp. \texttt{@a2}) can be sent over the xDSL link (resp. the \ac{LTE} interface). If the HCPE sends a packet whose source address is \texttt{@a2} on the xDSL link, it will be discarded by the provider. 

Given that two different addresses are assigned to the HCPE, a simple mechanism would be to assign these two addresses to the hosts attached to the HCPE. IPv6 has been designed with this possibility in mind and all IPv6 implementations can associate several IPv6 addresses to a single interface. With IPv4, this is more difficult as DHCP allocates a single address to each host. However, having several addresses per host would not solve our problem.
When a host has several addresses, it selects one of them every time it needs to establish a flow. If a host selects the address assigned by the xDSL network for the first packet of a flow, then the entire flow will be transported over the xDSL network. This implies that a given flow will have no way to utilize the two access links simultaneously. Furthermore, if one of the access links fails, then all the flows that were using this access link are broken, even if the other access link is still available. Another issue for network operators is that hosts can autonomously select their access link by selecting their source address. Many network operators view hybrid access networks as a solution to complement the xDSL network with the 4G network and they would like to use the 4G network only when the capacity available on the xDSL network is fully used. If the hosts autonomously decide the access link that they use, it becomes difficult to implement such policies.

Given the importance of hybrid access networks, several solutions have been explored by the \ac{BBF} and the \ac{IETF} to solve the above-mentioned problems. Three of them are defined in the recent BBF WT-378 specification \cite{bbf378}. They focus on three key requirements: $(i)$ supporting per-packet load balancing to be able to efficiently utilize two different access links for a single TCP connection, $(ii)$ assigning a single address to the end-hosts attached to the HCPE and $(iii)$ supporting Network Ingress Filtering. They all rely on the utilization of a special network function, called the \ac{HAG}, that is deployed by the network operator.  Figure~\ref{fig:architecture} describes the architecture of such Hybrid Access Networks. We assume a HCPE that is attached to both a DSL and an \ac{LTE} network. It has received one address from each network (\texttt{@cpe1} was allocated by the DSL network and \texttt{@cpe2} was allocated by the \ac{LTE} network). A single address is assigned to each connected host. We also assume that the \ac{HAG} is reachable via two addresses (\texttt{@h1} via the DSL network and \texttt{@h2} via the \ac{LTE} network).

\subsection{GRE-based L3 overlay tunneling}
The first realization of this architecture relies on \ac{GRE} Tunnels \cite{bbf378,rfc8157}. This solution can be summarized as follows. The HCPE is reachable via two different addresses (\texttt{@cpe1} and \texttt{@cpe2} in Figure~\ref{fig:architecture}), but those addresses are not directly exposed to the hosts that are attached to it. The HCPE has a third address that it assigns to its attached host. The \texttt{@cpe1} and \texttt{@cpe2} addresses assigned to the HCPE are only used by to create tunnels towards the HAG. When a HCPE boots, it creates these two tunnels and advertises its third address to the HAG. When a host initiates a TCP connection, it sends the first packet to the HCPE. The HCPE encapsulates it in one of its \ac{GRE} tunnels and sends it to the HAG. This encapsulation is required to support Network Ingress Filtering. The HAG decapsulates the packet and forwards it to its final destination. The server response reaches the HAG that maps it to its associated HCPE and encapsulates it in one of the tunnels that reaches the HCPE. The HCPE decapsulates the packet and forwards it to the client host. Both the HAG and the HCPE can use per-packet load-balancing to distribute the load over the two access networks. The solution specified in BBF WT-378 \cite{bbf378} includes several extensions to the basic \ac{GRE} encapsulation. First, sequence numbers are added to each encapsulated packet. These sequence numbers are used on the HCPE and the HAG to reorder the packets that are received over the xDSL and the 4G networks since those have different delays. This reordering ensures that all the packets that belong to a single TCP connection are delivered in-sequence, which prevents the TCP collapse that we described earlier. Furthermore, the HCPE and the HAG regularly measure the performance of the two tunnels to dynamically determine the fraction of the packets that must be sent over each tunnel. Additional details are provided in \cite{rfc8157}. Somme lessons learned from the deployment of Tunnel-based hybrid access networks have been discussed in \ac{IETF} presentations \cite{Leymann}

\subsection{L4 Multipath using MPTCP}
The two other standardized architectures rely on Multipath TCP \cite{rfc6824}. Multipath TCP is a recent TCP extension that enables hosts to use different paths to exchange packets belonging to a single TCP connection. Multipath TCP includes several mechanisms that enable hosts to efficiently utilize different networks \cite{nsdi12}. A Multipath TCP connection starts with a three-way handshake like a regular TCP connection.
To inform the server of its willingness to use Multipath TCP, the client inserts in the first \texttt{SYN} packet the \texttt{MP\_CAPABLE} option. This option contains a 64 bits key that is associated to this specific connection. The server replies with a \texttt{SYN+ACK} that also contains an \texttt{MP\_CAPABLE} option. The client replies with an \texttt{ACK} packet that also contains the \texttt{MP\_CAPABLE} option \cite{rfc6824}. At this point, the Multipath TCP connection is established over the path used by the \texttt{SYN} packets and both the client and the server can send data. Multipath TCP supports an \texttt{ADD\_ADDR} option that enables a host to advertise its other addresses to the remote host. To use another path, the client or the server must create a TCP subflow along this path. This is done by sending a \texttt{SYN} packet that contains the \texttt{MP\_JOIN} option. Thanks to the contents of this option, the server can associate the subflow establishment attempt to an existing Multipath TCP connection and authenticate it during the three-way handshake that creates the subflow. Once the subflow has been established, data can flow over any of the available paths. Multipath TCP uses coupled congestion schemes that enable it to react efficiently to congestion on the different paths \cite{nsdi12}. Furthermore, Multipath TCP uses its own sequence numbers to reorder the packets sent over different paths at the receiver but also to allow packet reinjection from one path to another one, for example upon timeout on the initially attempted path \cite{nsdi12}.

\begin{figure}[ht]
     \centering
     \includegraphics[width=0.45\textwidth]{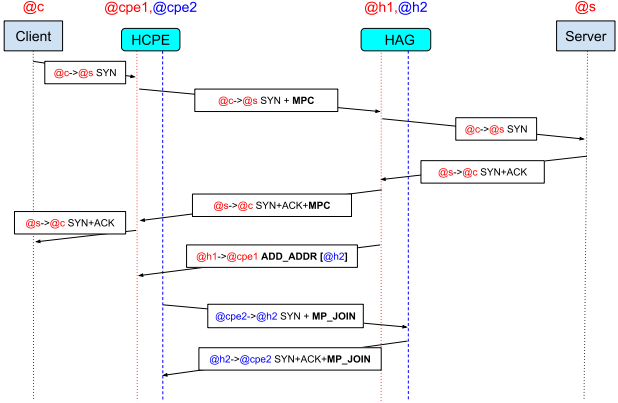}
     \caption{The implicit Multipath TCP deployment mode}
     \label{fig:transparent}
 \end{figure}

With Multipath TCP, a endhost could efficiently use the different paths that exist in hybrid networks. Unfortunately, although several implementations of Multipath TCP exist \cite{BonaventureSeo}, they are not used on popular endhosts nor on servers. %The largest deployment of Multipath TCP is on Apple's iPhones that use it to support the Siri voice recognition application \cite{BonaventureSeo}. Multipath TCP enables Siri to seamlessly switch from Wi-Fi to cellular or the opposite. However, besides Apple for specific applications, most end-hosts do not currently support Multipath TCP. 
To still benefit from the unique capabilities of Multipath TCP, the \ac{BBF} has decided to rely on Multipath TCP proxies. A Multipath TCP proxy is a network function that converts Multipath TCP connections into TCP connections and vice versa. It can be installed on HCPEs and HAGs. In the implict deployment \cite{bbf378}, the HAG is located on the forwarding path between the HCPE and the Internet on the xDSL network (Figure~\ref{fig:architecture}). Figure~\ref{fig:transparent} provides a sequence diagram of the establishment of a TCP connection using this implicit deployment. When a host attached to a HCPE initiates a connection, it sends the first packet to the HCPE. The Multipath TCP proxy running on the HCPE intercepts the TCP connection establishment attempt (\texttt{SYN} packet) and replaces it with a Multipath TCP connection establishment attempt (\texttt{SYN+MPC} packet). This packet is forwarded along the DSL network and the HAG intercepts it. The HAG performs the reverse operation and forwards a connection establishment attempt towards the final destination. The destination replies and confirms the connection establishment. This confirmation is intercepted by the HAG that in reaction confirms the establishment of the Multipath TCP connection to the HCPE which eventually confirms the establishment of the connection to the host. At this point, the connection between the host and the server is composed of three parts: $(i)$ a (regular) TCP connection between the host and the HCPE, $(ii)$ a (Multipath) TCP connection between the HCPE and the HAG and $(iii)$ a (regular) TCP connection between the HAG and the server. At that time, the HAG can advertise its address on the LTE network (i.e. address \texttt{@h2} on Figure~\ref{fig:architecture}) by using the \texttt{ADD\_ADDR} option. With this information, the HCPE can initiate a second subflow over the \ac{LTE} interface (\texttt{SYN+MP\_JOIN} packets exchanged with the HAG). % to enable the utilization of this second path. 
It should be noted that as both the HCPE and the HAG transparently intercept the TCP connections, they do not need to implement any Network Address Translation function. %The server sees the address of the host and not an address assigned to the HAG. 
This deployment supports per-packet load balancing as Multipath TCP continuously measures the packet losses and delays over the xDSL and the \ac{LTE} networks and dynamically adjusts the load over the two paths thanks to its congestion control scheme.

 \begin{figure}[ht]
     \centering
     \includegraphics[width=0.45\textwidth]{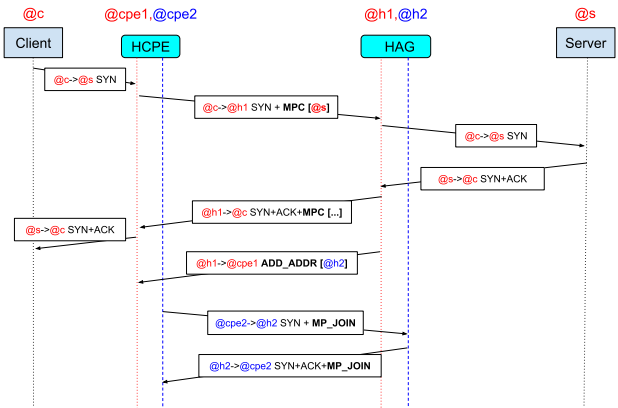}
     \caption{The explicit Multipath TCP deployment mode}
     \label{fig:converter}
 \end{figure}

The third approach also leverages Multipath TCP, but the HAG does not need to transparently intercept the Multipath TCP connections proxied by the HCPE. The packet flow is illustrated in Figure~\ref{fig:converter}. The HCPE still includes a transparent proxy that proxies the TCP connections established by the local hosts. The main difference with the previous deployment is that the HAG is directly reachable through an IP address over the xDSL network (\texttt{@h1} in Figure~\ref{fig:architecture}). Upon reception of a connection establishment packet from a local host towards a remote server, the HCPE sends a Multipath TCP connection establishment packet towards the address of its HAG and places inside the payload of this packet the address of the remote server (i.e. \texttt{@s}). This leverages the 0-RTT Convert protocol whose specification is being finalized within the \ac{IETF} \cite{draft-converters}. The HAG terminates the Multipath TCP connection and immediately sends a connection establishment packet towards the remote server. As in the previous solution, the HCPE can also create a subflows over the \ac{LTE} network to utilize the two access networks. The connection between the client and the server is composed of three connections that are glued together by the transparent proxy running on the HCPE and the 0-RTT convert protocol that is running on the HAG. This is illustrated in Figure \ref{fig:converter}.

 %\begin{figure}
 %    \centering
 %    \includegraphics[width=0.45\textwidth]{figures/converter.eps}
 %    \caption{The explicit Multipath TCP deployment mode}
 %    \label{fig:converter}
 %\end{figure}

%Multipath TCP (MPTCP) was accepted as an experimental standard by the Internet Engineering Task Force (IETF) in January 2013 \cite{rfc6824}. It was then used in other RFCs and in the Broadband Forum (BBF) technical report TR-348.
%\subsection{IETF}
%In addition to the RFC6824 which is the first standardization work of this extension to the TCP protocol called MPTCP, other drafts were submitted to standardized how MPTCP could be used in access networks.
%\subsection{Broadband Forum}
%The main technical report for hybrid access networks at the BBF is TR-348 \cite{bbf348}.

\section{Implementation and deployment experience}\label{section:implem}

As explained in the previous section, the architecture proposed by the \ac{BBF} for Hybrid Access Networks relies on two components: $(i)$ a Hybrid Aggregation Gateway and $(ii)$ a Hybrid-CPE. We describe in this section our experience in implementing and deploying those two components and evaluate their performance.

\subsection{Hybrid CPEs}

A HCPE is defined as a CPE which is attached to both an xDSL and a 4G network.
%Several types of HCPEs have been deployed. Some operators use a single box with both xDSL and 4G interfaces. Others prefer to reuse their existing xDSL CPE with minor modifications and attachMost CPEs are based on Linux. %A HAG typically serves a large number of active customers. 
Most of the deployed xDSL and 4G CPEs use Linux. We have extended several of these with the open-source Multipath TCP implementation in the Linux kernel \cite{mptcplinux}. This stack completely supports the Multipath TCP protocol and is considered to be its reference implementation. We extend it with a transparent proxy that efficiently moves data from one Multipath TCP connection to one TCP connection and vice-versa. These data transfer operations are implemented in the Linux kernel for performance reasons \cite{Detal}.

One HCPE typically serves only a few clients in a home network, connected through Wi-Fi or Ethernet. Performances are still important because the CPU of the CPE has other services to run besides the MPTCP proxy. Some deployments rely on a HCPE that supports both xDSL and \ac{LTE} while others use a regular xDSL CPE that is attached over a Gigabit Ethernet interface to an \ac{LTE} CPE. This simplifies the deployment as existing xDSL CPEs can be reused.

\begin{figure}[ht]
    \center
    \includegraphics[width=0.45\textwidth]{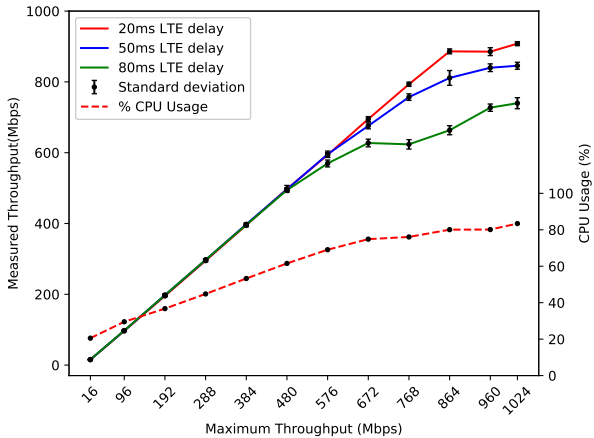}
    \caption{Total throughput and CPU usage on Linksys router}
    \label{fig:linksys}
\end{figure}

The maximum throughput that a HCPE can sustain mainly depends on the performance of its CPU. Older CPEs can reach a throughput of 100 Mbps while more recent ones such as most 4G gateways can reach much higher throughput. As an example, we evaluated the performance of a simple off-the-shelf CPE: the Linksys WRT1200AC. This CPE is equipped with a Marvell Armada 385 CPU clocked at 1.33Ghz. We attach a client running \texttt{iPerf} to this CPE and emulate the two access networks by using Gigabit Ethernet links, with different delays. A 20ms delay on one link to emulate the DSL network, and between 20ms and 80ms to emulate the LTE network. The difference of delay between the two link will have an impact on the reordering of the packets.
We use \texttt{iperf} to generate ten downlink TCP connections that transfer data during 30 seconds. The baseline throughput for the setup on a Gigabit Ethernet link with 20ms delay is 936Mbps, with a total CPU usage of 70\%. Figure~\ref{fig:linksys} plots, in plain lines, the \texttt{iperf} throughput measured on the client in function of the bandwidth available on both links, with a symmetric split 50\% on each. The values of the maximum bandwidth on the X axis are ranging from 1 Mbyte/sec to 128 Mbytes/sec, converted in Mbps, and have a maximum at 2x512 Mbps, because the client is connected to the CPE with a Gigabit Ethernet link. Measurements are average over 5 iterations, and the standard deviation is indicated in black. Each line corresponds to a different delay on the emulated LTE link, while the delay for the emulated DSL is always 20ms. It shows that the delay difference between the two links has no influence for throughputs below 576Mbps, but the aggregated speed decreases with an increasing difference.
It also plots, in dashed line, the average usage of one of the CPU in function of the available bandwidth. We can see that the CPU usage increases with the aggregated throughput, but that a retail CPE can achieve up to 900Mbps by combining two 512Mbps links. 

%A HCPE also needs at least three network interfaces, including the two facing the HAG on the xDSL and \ac{LTE} networks. The third one is to connect to clients in the home network.

The network operators who have contributed to the standardization of hybrid access networks \cite{bbf378,bbf348} have insisted on the ability to support specific policies such as specifying the ports or address blocks which can benefit from aggregation, but also ensuring that the \ac{LTE} network is only used when the DSL one is saturated. One key HCPE function is to decide when to create an \ac{LTE} subflow based $(i)$ on operator objective to lower its costs and $(ii)$ user experience optimization, which is best when both networks are used. For subscribers, the experience depends on the available bandwidth that is the combination of bandwidth from the xDSL and \ac{LTE} networks. For the operator, the optimal business objective is to first use bandwidth from the least costly network (xDSL in the case considered here), and then use the minimum necessary bandwidth on the other network.
The trade-off between those apparently mutually exclusive objectives is achieved by the overflow concept. The HCPE always starts to use the xDSL network, until the available DSL capacity is fully used, and then overflows on the \ac{LTE} network. The overflow mode leverages two features of Multipath TCP \cite{mptcplinux,schedulers}. The HCPE continuously measures the load of the DSL link. If the average load is low, it does not attempt to create subflows over the \ac{LTE} network. If the average load is above a configured threshold (e.g. 80\% of the DSL link capacity), then it automatically opens one subflow over the \ac{LTE} network for each proxied TCP connection. We then rely on a specialized packet scheduler to preferentially use the xDSL network. Multipath TCP includes a flexible packet scheduler \cite{schedulers} that selects the subflow that is used to transmit each packet.

\begin{figure}[ht]
    \center
    \includegraphics[width=0.5\textwidth]{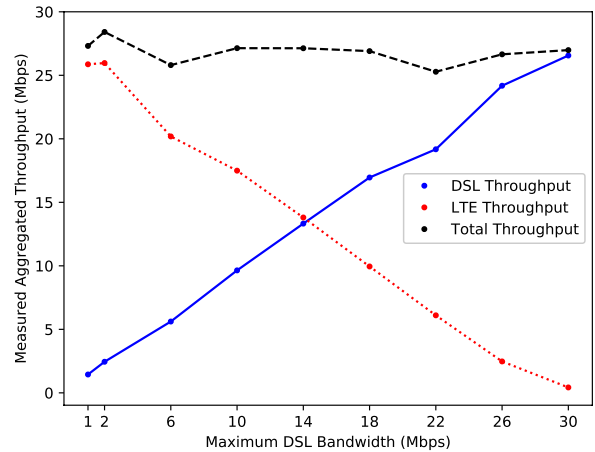}
    \caption{Thanks to the overflow mode, the HCPE first saturates the xDSL network before using the \ac{LTE} one.}
    \label{fig:RatioDSLLTE}
\end{figure}

This overflow is illustrated on Figure~\ref{fig:RatioDSLLTE} where a hybrid access is simulated with a CPE and a HAG acting as MPTCP proxies (as illustrated on Figure~\ref{fig:architecture}), and a client host running \texttt{iPerf}, with the two access networks simulated as Ethernet links between the HCPE and the HAG. We use \texttt{iperf} to generate a single downlink TCP connection that operates at 30 Mbps during 15 seconds. We consider the capacity of the \ac{LTE} link to be unlimited. Figure~\ref{fig:RatioDSLLTE} plots the measured throughput on the DSL link (blue line), \ac{LTE} link (red line) and aggregated (yellow line), averaged on the duration of the \texttt{iperf} test, in function of the theoretical capacity of the DSL link. As expected, the DSL link is at maximum capacity with a measured throughput following the theoretical capacity on the link, and the measured total aggregated throughput is around 30 Mbps. We observe that the measured throughput on the \ac{LTE} link corresponds to the difference between the 30 Mbps of the \texttt{iperf} and what is available on the DSL link. This clearly shows that the Hybrid Access Network utilizes both networks to achieve the required throughput, but utilizes the \ac{LTE} network only when the DSL one is saturated.

\subsection{Field Deployments}\label{section:deployments}
 
Several network operators have started to deploy Hybrid Access Networks using the Multipath TCP solutions described in this paper. Their main use case is to provide faster Internet access services in rural areas where the length of the twisted pair lines does not allow the xDSL technologies to reach more than a few Mbps. The customer surveys that they have conducted indicate that these Hybrid Access Networks improve the quality of experience perceived by the end-users. 

Several deployment models can be used for Hybrid Access Gateways. Ideally, they should be installed at locations where both the fixed and the cellular networks are present. Some operators deploy a small number of high capacity HAGs in their backbone. Others have a more decentralised network where both the fixed and the cellular network coexist at multiple locations. In this case, they deploy HAGs at all these locations. Many operators opt to deploy HAGs running on virtual machines to ease their management and provisioning.

\begin{figure}[ht]
    \center
    \includegraphics[width=0.45\textwidth]{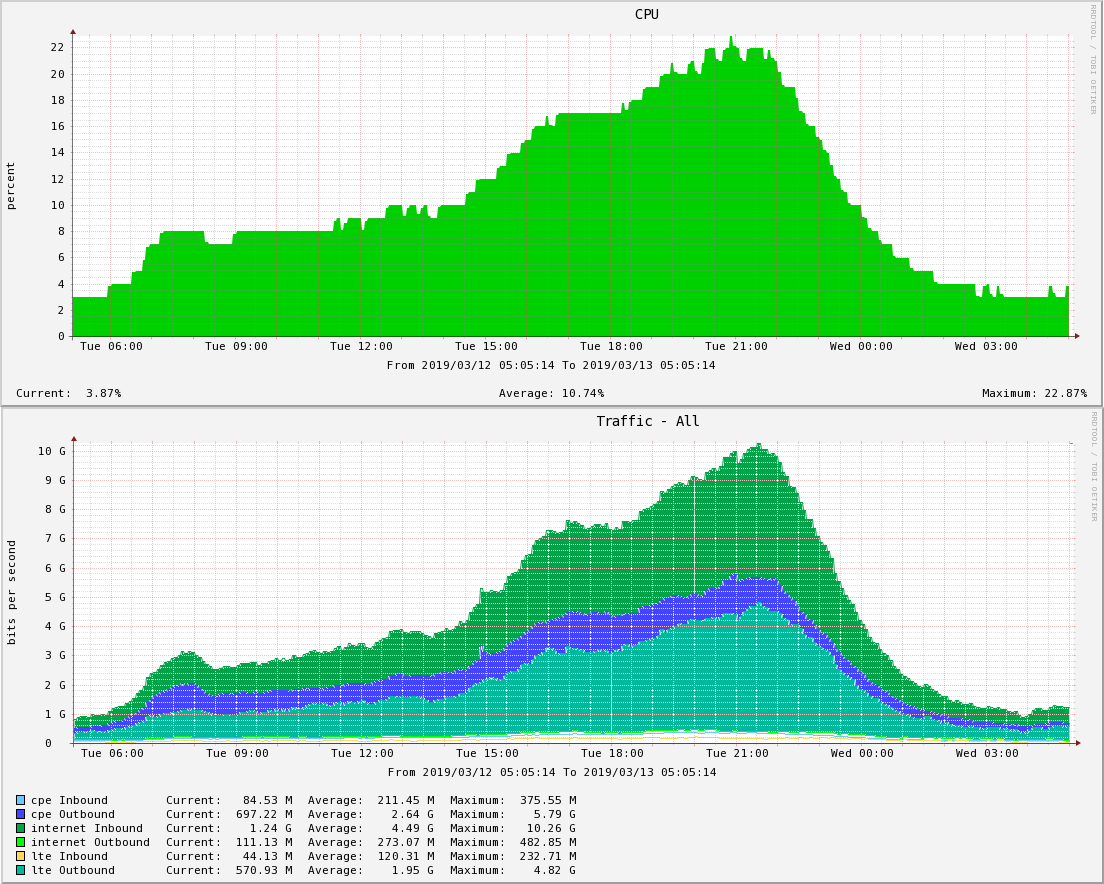}
    \caption{CPU utilisation (top) and bandwidth usage of a HAG serving 10k customers in a European network}
    \label{fig:kpn}
\end{figure}

As an illustration, Figure~\ref{fig:kpn} provides some of the statistics collected on one HAG deployed in a European network. This HAG uses the implicit mode and serves about 10k homes. This HAG is running 
on a %Server: HPE DL380 Gen10 
server equipped with Intel Xeon-Platinum 8180 (2.5GHz/28-core/205W) and four Mellanox ConnectX-4 10Gbps interfaces. The four interfaces are combined in a 40 Gbps \ac{LAG} and divided in three \ac{VLANs}: one attached to the DSL network, one to the LTE network and one to the Internet through a backbone router. 
%Lx based
%%To be adapted to new figures
%The first interface (top of Figure~\ref{fig:kpn}) is attached to the DSL network. The second (middle of Figure~\ref{fig:kpn}) is attached to the \ac{LTE} network. The third and fourth interfaces (bottom of Figure~\ref{fig:kpn}) are combined as a 20 Gbps LAG and attached to the \ac{PE} router that is connected to the Internet.

Figure~\ref{fig:kpn} confirms in a real deployment several of the points that we mentioned earlier. The statistics were collected on a normal week day. The peak hours are during the evening and most of the users download data. The top part of Figure~\ref{fig:kpn} provides the evolution of the CPU load of the server. It increases slowly with the traffic, but reaches less than 25\% on this server while serving a bit more than 10 Gbps of Internet traffic. During the peak hours, there were almost 7.5k active users and the HAG maintains on average 180k Multipath TCP connections. During the peak hours, the HCPEs establish up to 6100 MPTCP subflows per second through this HAG. 
The bottom part of Figure~\ref{fig:kpn} shows the distribution of the traffic over the different interfaces. The HAG receives up to 10.26 Gbps from the Internet and forwards up to 5.79 Gbps over the DSL network and 4.82 Gbps over the LTE network. The distribution between the DSL and the LTE network depends on the profiles of the users and the maximum bandwidth of the served DSL links and the bandwidth cap over the \ac{LTE} connections. We observed in many deployments that the DSL and LTE loads were similar although the LTE connections have a higher bandwidth than the DSL ones.%An operator can tune those parameters to provide different types of services to different types of users. For example, it is possible to boost the download for home users while using the \ac{LTE} network to boost the upload for teleworkers.

\section{Conclusion}

During the last few years, Hybrid Access Networks have matured. The architecture initially proposed by the \ac{BBF} has been realized by leveraging new \ac{IETF} protocols. We have described the key principles behind the tunnel-based and the Multipath TCP-based solutions. The first deployments combine DSL and \ac{LTE} to provide higher bandwidth services in rural areas where the length of the copper pairs does not enable the deployment of faster DSL services such as VDSL2. Our measurements indicate that the Multipath TCP-based solutions, which can be deployed as a software extensions on CPE routers and on commodity x86 servers, enable to use the \ac{LTE} network only once the DSL link becomes saturated while still supporting high bandwidth services. 

The ongoing deployments demonstrate that network operators can optimize their assets by combining different heterogeneous access networks to provide faster, cheaper or more resilient services to their customers. Although the first Hybrid Access Networks combined xDSL and LTE, the technology is generic and can be applied to any type of network. Recently, the 3GPP has decided to also reuse the 0-RTT convert protocol to support the Access Traffic Steering, Switch and Splitting (ATSSS) that will enable 5G networks to combine 5G with other technologies such as Wi-Fi or satellite. %  also open new opportunities to develop more advanced services that better leverage the unique characteristics of the different networks that are combined together. %Although the initial deployments combine DSL and \ac{LTE}, any pair of access networks can be combined provided that they support IP. 

% \appendices
% \section{}

\section*{Acknowledgments}
We would like to thank the Tessares employees who have participated in the development of the hybrid access network solutions described in this paper as well as Bart Peirens, Mohammed Boucadair and Guiu Fabregas for their contributions to the \ac{IETF} and \ac{BBF} specifications.

\bibliographystyle{plain}

\begin{thebibliography}{1}
% maximum 15

\bibitem{ECreport}
IHS Markit Ltd., Point Topic, \emph{Broadband Coverage in Europe}, 2018, doi:10.2759/358688, [Online]. Available: \url{http://ec.europa.eu/newsroom/dae/document.cfm?doc_id=52968}. [Accessed: 26-Mar-2019].
%http://ec.europa.eu/newsroom/dae/document.cfm?doc_id=52968

\bibitem{fcc}
FCC, \emph{2018 Broadband Deployment Report}, [online], https://www.fcc.gov/reports-research/reports/broadband-progress-reports/2018-broadband-deployment-report [Accessed: 01-Apr-2019]

\bibitem{bbf348}
Broadband Forum, \emph{TR-348 Hybrid Access Broadband Network Architecture}, Issue 1.\hskip 1em plus
  0.5em minus 0.4em\relax August 2016.
  
\bibitem{bbf378}
Broadband Forum, \emph{Nodal Requirements for Hybrid Access Broadband Networks}, WT-378, 2019

\bibitem{rfc6824}
A. Ford, C. Raiciu, M. Handley, O. Bonaventure, \emph{TCP Extensions for Multipath Operation with Multiple Addresses}, RFC6824, January 2013

\bibitem{rfc8157}
N. Leymann, C. Heidemann, M. Zhang, B. Sarikaya, M. Cullen, \emph{Huawei's GRE Tunnel Bonding Protocol}, RFC8157, May 2017

\bibitem{draft-converters}
O. Bonaventure, M. Boucadair, S. Gundavelli, S. Seo, B. Hesmans, \emph{0-RTT TCP Convert Protocol}, Internet draft, draft-ietf-tcpm-converters-06, work in progress, March 2019

\bibitem{rfc2827}
P. Ferguson, D. Senie, \emph{Network Ingress Filtering: Defeating Denial of Service Attacks which employ IP Source Address Spoofing}, RFC2827, May 2000

%\bibitem{rfc6724}
%D. Tharler, R. Draves, A. Masumoto, T. Chown, \emph{Default Address Selection for Internet Protocol Version 6 (IPv6)}, RFC6724, Sept.2012

\bibitem{Leymann}
N. Leymann, \emph{Hybrid Access deployment @ DT}, Banana BoF, IETF95, Buenos Aires, April 2016
%https://www.ietf.org/mail-archive/web/banana/current/pdfznR7ujMPhP.pdf

\bibitem{nsdi12}
C. Raiciu, et al. \emph{How hard can it be? designing and implementing a deployable Multipath TCP} Proceedings of the 9th USENIX conference on Networked Systems Design and Implementation. USENIX Association, 2012.

\bibitem{BonaventureSeo}
O. Bonaventure, S. Seo, \emph{Multipath TCP Deployments}, IETF Journal, Nov. 2016

\bibitem{mptcplinux}
C. Paasch, S. Barr\'e et al., \emph{Multipath TCP in the Linux kernel}, [Online]. Available: \url{https://www.multipath-tcp.org}. [Accessed: 26- Mar- 2019].

\bibitem{Detal}
G. Detal, C. Paasch, O. Bonaventure, \emph{Multipath in the Middle(Box)}, HotMiddlebox'13, Santa Barbara, 2013

%\bibitem{boucadair}
%M. Boucadair, C. Jacquenet, T. Reddy, \emph{DHCP Options for 0-RTT TCP Converters},Internet draft, draft-boucadair-tcpm-dhc-converter-01, work in progress, Oct. 2018

%\bibitem{mptcplinux}
%C. Paasch, S. Barre et al., \emph{Multipath TCP in the Linux kernel}, available from \url{https://www.multipath-tcp.org}

\bibitem{schedulers}
C. Paasch, S. Ferlin, O. Alay, O. Bonaventure, \emph{Experimental evaluation of Multipath TCP schedulers}. In Proceedings of the 2014 ACM SIGCOMM workshop on Capacity sharing workshop (pp. 27-32). ACM, 2014
















\end{thebibliography}


%\begin{IEEEbiographynophoto}{Benjamin Hesman} works as an R\&D engineer at Tessares and is currently finalising his Ph.D. on Multipath TCP at UCLouvain. \end{IEEEbiographynophoto}

%\begin{IEEEbiographynophoto}{Nicolas Keukeleire}
%is Product Manager at Tessares for the hybrid access software %solution. He has graduated in Electrical Engineering from the Universit\'e Libre de Bruxelles and worked at Belgacom International Carrier Services.

%\begin{IEEEbiographynophoto}{Olivier Bonaventure}
%is Professor at UCLouvain where he leads the IP Networking Lab. He has actively contributed to the design, implementation and deployment of several Internet protocols including LISP, Multipath TCP, IPv6 Segment Routing, ... He currently serves as Editor for SIGCOMM's Computer Communication Review. He wrote the \emph{Computer Networking: Principles, Protocols and Practice} open-source ebook that is used by various universities. He co-founded the Tessares spinoff that deploys Hybrid Access Networks using Multipath TCP.

\end{document}